\documentclass[journal=nalefd,manuscript=letter,layout=twocolumn]{achemso}

\usepackage[version=3]{mhchem} % Formula subscripts using \ce{}
\usepackage{lineno}
\usepackage{graphicx}% Include figure files
\usepackage{dcolumn}% Align table columns on decimal point
\usepackage{bm}% bold math
\usepackage{caption}

\author{Robert M. Pettit}
\altaffiliation{These authors contributed equally to this work.}
\email{robert@memq.tech}
\affiliation{memQ, Inc, Chicago, IL 60615, USA}

\author{Skylar Deckoff-Jones}
\altaffiliation{These authors contributed equally to this work.}
\email{skylar@memq.tech}
\affiliation{memQ, Inc, Chicago, IL 60615, USA}

\author{Angela Donis}
\affiliation{memQ, Inc, Chicago, IL 60615, USA}

\author{Ana Elias}
\affiliation{memQ, Inc, Chicago, IL 60615, USA}

\author{Jayson Briscoe}
\affiliation{The American Institute for Manufacturing Integrated Photonics (AIM Photonics), Albany, NY 12203, USA}

\author{Gerald Leake}
\affiliation{The American Institute for Manufacturing Integrated Photonics (AIM Photonics), Albany, NY 12203, USA}

\author{Daniel Coleman}
\affiliation{The American Institute for Manufacturing Integrated Photonics (AIM Photonics), Albany, NY 12203, USA}

\author{Michael Fanto}
\affiliation{Air Force Research Laboratory, Information Directorate, Rome, NY 13441, USA}

\author{Ananthesh Sundaresh}
\affiliation{memQ, Inc, Chicago, IL 60615, USA}

\author{Shobhit Gupta}
\affiliation{memQ, Inc, Chicago, IL 60615, USA}

\author{Manish Kumar Singh}
\affiliation{memQ, Inc, Chicago, IL 60615, USA}

\author{Sean E. Sullivan}
\affiliation{memQ, Inc, Chicago, IL 60615, USA}

\title{Monolithically Integrated C-Band Quantum Emitters on Foundry Silicon Photonics}

\keywords{Silicon Photonics, Single Photon Emitter, Erbium, Photonic Crystal Cavity, Quantum Networking, Purcell Enhancement}

\begin{document}

%\section{Abstract}
\begin{abstract}
Solid-state spin-based quantum systems have emerged as popular platforms for quantum networking applications due to their optical interfaces, their long-lived quantum memories, and their natural compatibility with semiconductor manufacturing. Photonic crystal cavities are often used to enhance radiative emission; however, fabrication of the necessary subwavelength cavities is typically limited to small batch electron beam lithography. In this work, we demonstrate high quality factor, small mode volume nanobeam cavities fabricated on a scalable silicon photonic foundry platform. The foundry fabricated cavities are then interfaced with single erbium ions through backend deposition of TiO$_2$ thin films lightly doped with erbium. Single ion lifetime measurements indicate Purcell enhancement up to about 500, thereby demonstrating a route toward manufacturable deterministic single photon sources in the telecom C-band.
\end{abstract}

%\section{Introduction}
Networking of different quantum devices using optical fiber and flying qubits is critical for the realization of distributed quantum computing, quantum sensor networks, and secure quantum communication \cite{wehnerQuantumInternetVision2018, weiRealWorldQuantumNetworks2022, cuomoDistributedQuantumComputing2020, pettitPerspectivePathwayScalable2023}. Solid-state spin qubit platforms are particularly appealing for these applications as their spin enables long lived quantum memories and they possess natural optical transitions to coherently convert to photonic qubits \cite{awschalomQuantumTechnologiesOptically2018}. Rare earth ions (REIs) in particular have the additional advantage of long spin coherence and, in the case of trivalent erbium (Er$^{3+}$), desirable C-band emission for low-loss long-distance fiber propagation. While there have been multiple pioneering works demonstrating isolation and coherent control of individual Er$^{3+}$ ions, these experiments focused on doped bulk crystals (several mm in size) that are not compatible with semiconductor foundry processing \cite{dibosAtomicSourceSingle2018, craiciuNanophotonicQuantumStorage2019, kindemControlSingleshotReadout2020, rahaOpticalQuantumNondemolition2020}. Additionally, doping of thin films such as lithium niobate has been explored as a route toward photonic integration of single REIs using smart-cut techniques \cite{yang_controlling_2023}. Recently it has been shown that REIs can be grown in oxide thin films for monolithic integration with silicon photonics platforms \cite{singhEpitaxialErdopedY$_2$O$_3$2020, dibosPurcellEnhancementErbium2022, singhOpticalMicrostructuralStudies2024, jiNanocavityMediatedPurcellEnhancement2024}. Titanium dioxide (TiO$_2$) in particular has been demonstrated as a good host material for Er$^{3+}$ ions and is compatible with complementary metal-oxide-semiconductor (CMOS) processing \cite{phenicieNarrowOpticalLine2019}. These films can then be deposited onto silicon-on-insulator (SOI) and patterned into nanophotonic cavities and circuits, where silicon photonic waveguides enable routing and switching and REI optical emission rates are enhanced through the Purcell effect via evanescent coupling to integrated cavities. Using this approach, individual Er$^{3+}$ ions were recently isolated in REI doped oxides on SOI where high Purcell factor-enabling cavities were fabricated through electron beam lithography (EBL) \cite{jiIsolationIndividualEr2024}. Although electron beam lithography is suitable for fabricating hundreds of devices on a chip, it does not offer the wafer-scale throughput or process control required for mass manufacturing. In contrast, advanced 300 mm CMOS foundries provide a clear path to reproducible, production-scale manufacturing of these devices.

Subwavelength integrated photonic structures, such as high-Q and small mode volume photonic crystals, are key components to optically interface with solid-state qubits. These structures may also benefit active photonic components such as lasers and modulators due to the devices reduced footprint \cite{zhangPhotonicCrystalNanobeam2010, gongNanobeamPhotonicCrystal2010, zhongUltralowpowerConsumptionSilicon2024}. Traditionally, these devices have been difficult to mass-produce, relying on low-volume methods like EBL. Meanwhile, many commercial integrated photonic foundries, which use the same facilities to fabricate advanced CMOS technology nodes, can now achieve feature sizes below 100 nm. Crucially, the ability to fabricate subwavelength features in the optical layer unlocks wafer-scale and repeatable manufacturing of a new variety of performative and ultra-compact filters, couplers, and cavities \cite{chebenSubwavelengthIntegratedPhotonics2018}. There have been several demonstrations of high-Q photonic crystal-based cavities fabricated on foundry silicon photonics. However, devices are typically not designed with the small mode volumes necessary for Purcell enhancement of quantum emitters \cite{dodaneFullyEmbeddedPhotonic2018, arnoldDeepSubwavelengthSlotted2024}. Recently, a demonstration of small mode volume and high-Q suspended 2D photonic crystal cavities was achieved over multiple process optimization runs \cite{panuski_full_2022}. Fabrication-robust photonic crystals have also been designed for the particular purpose of interfacing to solid-state qubits, although they have yet to be fabricated on a foundry process \cite{olthausOptimalPhotonicCrystal2020, abulnagaDesignFabricationRobust2024}. In addition to refinements of critical dimension size via optimized optical lithography, many foundry platforms now offer trenches through the top cladding to the photonic layers meant for evanescent sensing, and these trenches also present the opportunity for backend integration of qubits onto the waveguides.

In this work, we fabricate silicon photonic crystal nanobeam cavities using AIM Photonics’ 300 mm silicon photonic platform and characterize the cavities’ local and wafer scale performance variation. In the backend, Er$^{3+}$:TiO$_2$ thin films are deposited on top of exposed cavities through sensing trenches, providing a facile route for the integration of novel materials with a foundry photonic platform. The impact of the thin film on cavity performance is then measured. We further use the nanobeam cavities to enhance optical emission from single Er$^{3+}$ ions through the Purcell effect \cite{purcellSpontaneousEmissionProbabilities1995}, creating single photon sources in the C-band on a scalable foundry 300 mm wafer platform. In addition, we also fabricate photonic crystal nanobeam cavities in the silicon nitride layer and Bragg filters for pump rejection \cite{oserCoherencyBrokenBraggFilters2019}, and their performance is presented in the Supporting Information.

%\section{Results and Discussion}
The devices were fabricated on AIM Photonics' QFlex 300mm PIC platform. The AIM Photonics MPW Service leverages the Albany Nanotech Complex’s state-of-the-art 300 mm R\&D fab, which also develops sub-7 nm CMOS technology. This industrial-grade environment provides precise process control, enabling rapid production and reliable device performance. The platform uses 220 nm silicon-on-insulator (SOI) wafers with a thick buried oxide for substrate isolation. Patterns are defined using 193 nm immersion lithography, which offers critical dimension control needed to realize the presented photonic crystal cavities. Advanced reactive ion etching (RIE) with subsequent chemical treatments ensures the creation of smooth, vertical sidewalls. The structures are then clad in a low-loss pTEOS dielectric, applied with a non-conformal deposition technique to minimize defects. 

\begin{figure*}
\centering
\includegraphics[width=6in]{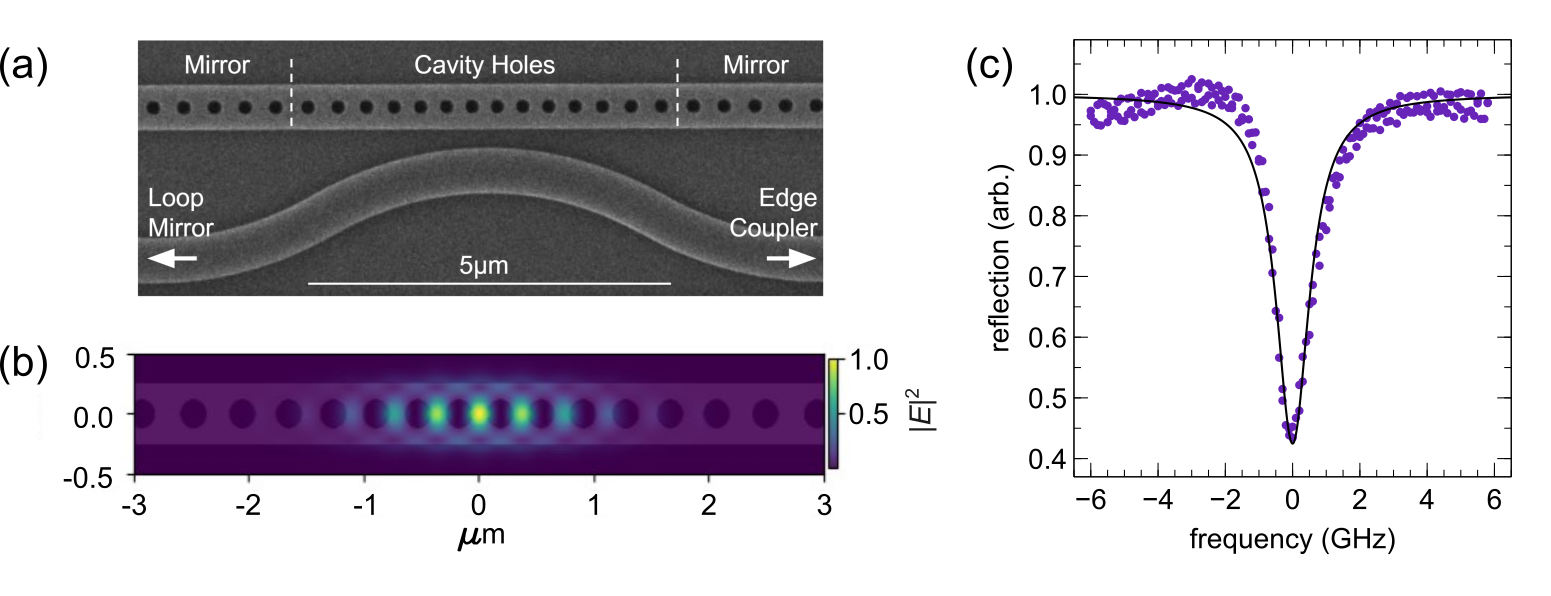} %Fig1_Si_cavity_v2.png
\caption{(a) SEM image of fabricated photonic crystal nanobeam cavity. (b) Simulated optical field intensity in the silicon nanobeam cavity. (c) Reflection spectrum of the silicon nanobeam cavity fit with a Lorentzian line shape to determine the quality factor. }
\label{fig:cavity1}
\end{figure*}

The photonic crystal nanobeam cavities comprise two photonic crystal mirrors defined by holes etched into a silicon waveguide to create a photonic bandgap in the telecom C-band as shown in Fig \ref{fig:cavity1}a. In between the mirrors, the pitch of the holes are reduced parabolically to create a resonant midgap state\cite{quanDeterministicDesignWavelength2011}. The optimized simulated cavity achieved a quality factor of 780,000, with a resonance wavelength of 1537nm. Figure \ref{fig:cavity1}b shows the simulated field intensity of the resonant mode in the silicon cavity, which achieves a small mode volume $<$ 0.4 ($\lambda$/n)$^3$ due to the high index contrast between silicon and the silicon dioxide cladding. A bus waveguide is then run adjacent to the cavity, where the size of the gap between the bus waveguide and the cavity determines the cavity coupling. Device parameters such as hole size, waveguide width, and coupling gap are varied to optimize device coupling, resonant wavelength, and quality factor. One end of the coupling waveguide goes to an integrated Sagnac loop mirror, while the other end goes to an edge coupler. A Sagnac loop mirror was used to provide uniform broadband reflection compared to a Bragg mirror, and the edge couplers were chosen as a PDK component to provide broadband coupling off--chip with a coupling efficiency of 3~dB. The devices are characterized in a reflection configuration using a circulator. A reflection spectrum of a silicon cavity is shown in Fig. \ref{fig:cavity1}c with a fit to a Lorentzian line shape giving a quality factor of 158,046(30). The discrepancy between simulated and experimental performance is due to roughness induced scattering which can’t be easily simulated. 

More than 5,000 devices (per reticle) with varying design parameters were characterized using an automated probe station, enabling correlation between design parameters and performance. Fig. 2a shows the silicon cavities resonant wavelength as a function of cavity hole diameter while holding the cavity waveguide width constant at 450 nm. As the hole diameter is reduced, the resonant wavelength redshifts linearly, creating a convenient way to control the cavity's resonant wavelength. For a given hole size, there is still a relatively large distribution, with an interquartile range of $>$5~nm, because the cavities are extremely sensitive to fabrication variation. Fig. 2b shows the quality factor of cavities as a function of coupling gap size between the bus waveguide and the cavity while the hole diameter and waveguide width are held constant at 185 nm and 500 nm, respectively. The quality factors increase with coupling gap size as the higher Q cavities become closer to critically coupled. This suggests that the cavities were over-coupled and in future iterations larger coupling gaps should be targeted to achieve critically coupled cavities. There is a large variance in the device Q factor, which again stems from the extreme sensitivity of these small-mode volume cavities to nanoscale fabrication variations. The best performing devices were identified after characterizing a wide design sweep of hole diameter and coupling gap from the nominal optimized simulated design. The minimum quality factors remain similar as some cavities remain overcoupled or undercoupled for a given gap size. The largest measured quality factors were found for devices with the largest gap sizes where they were near critically coupled. The best performing silicon cavities achieved quality factors in excess of 150,000, however we note that devices of identical layout had large variation of Q-factor (~50,000-150,000). This can be attributed to the extreme sensitivity to fabrication variation in these small mode volume devices. This can potentially be improved in future device iterations by designing cavities where the mode is more confined in the silicon compared to the etched surfaces. Many of the device variations here included trenches through the oxide top cladding to the silicon nanobeam cavities, enabling back-end processing of other materials onto the cavities - CMOS compatible Er$^{3+}$:TiO$_2$ in this case \cite{jiIsolationIndividualEr2024,jiNanocavityMediatedPurcellEnhancement2024}. Device performance from different reticles across the wafer was also characterized, and showed similar performance with a slight radial dependence (see Supporting Information). 

\begin{figure*}
\centering\includegraphics[width=6in]{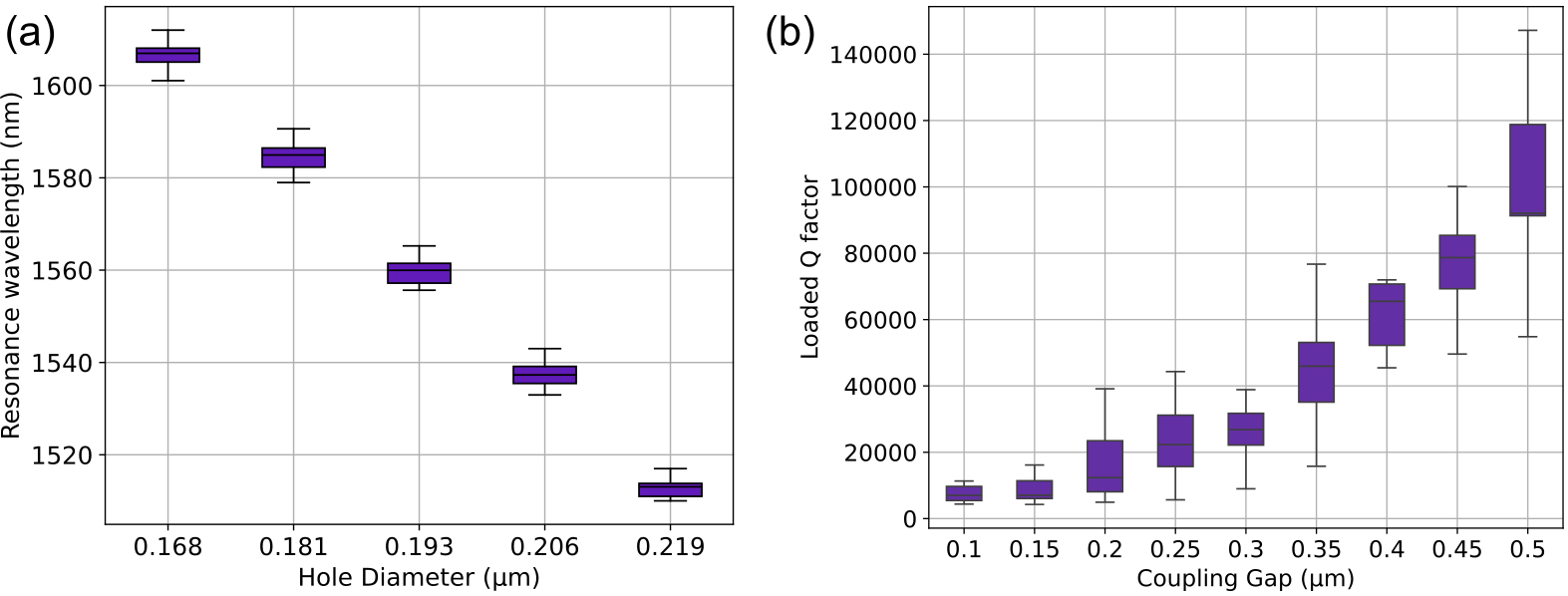} %Figure2_v2
\caption{(a) Box and whisker plot (interquartile range and +/- 1.5 quartiles) of the cavity resonance wavelength as a function of photonic crystal cavity hole diameter (b) Box and whisker plot of the cavities quality factor as a function of the gap between the bus waveguide and the cavity. Each box and whisker contains data from 15 devices.}
\end{figure*}

Upon receiving the wafers from AIM Photonics, Er$^{3+}$:TiO$_2$ films were grown on the photonic crystal through trenches down to the silicon optical layer on the PICs, as shown in Fig. 3a. Many silicon photonic foundries offer a trench to the photonic layer which is generally meant for sensing applications. Here we use it as a way to introduce the ``quantum layer'' post fabrication. This approach demonstrates a simple way to integrate solid-state quantum materials with commercial foundry integrated photonics without complex lithography or alignment. Recent demonstrations have shown high-quality Er$^{3+}$:TiO$_2$ thin films can also be grown using Atomic Layer Deposition (ALD), which is compatible with standard CMOS processing \cite{ji_nanocavity-mediated_2024}. 300mm ALD tools are used for growing high quality gate dielectrics and diffusion barriers and are already present in AIM Photonics' foundry line. Therefore, integrating Er$^{3+}$:TiO$_2$ into the front-end foundry process through ALD is a straightforward future step that would enhance film quality and scalability. 

The quantum layer consists of two 10 nm buffer layers of TiO$_2$ with a 1 nm-thick delta-doped layer of Er$^{3+}$ with a concentration of 2 ppm, yielding approximately 200 ions within a cavity mode volume. The film is polyphase, containing both rutile and anatase phases of TiO$_2$ and has an RMS surface roughness of 2.4 nm as measured by AFM. The films are grown with molecular beam deposition as reported elsewhere \cite{singhOpticalMicrostructuralStudies2024}.  Following deposition, the devices were remeasured, and it was found that the resonant wavelengths of the cavities uniformly red shifted by 30 nm, as shown in Fig. 3b. The distribution of the resonance wavelengths remained relatively unchanged. The quality factors of the cavities were also compared before and after the addition of the quantum layer, as shown in Fig. 3c. There was a slight reduction in the cavity Q-factor, which might be a result of the film's surface roughness or material absorption; nevertheless, several cavities still had cavity resonances with Q-factors in excess of 100,000. 

\begin{figure*}
\centering
\includegraphics[]{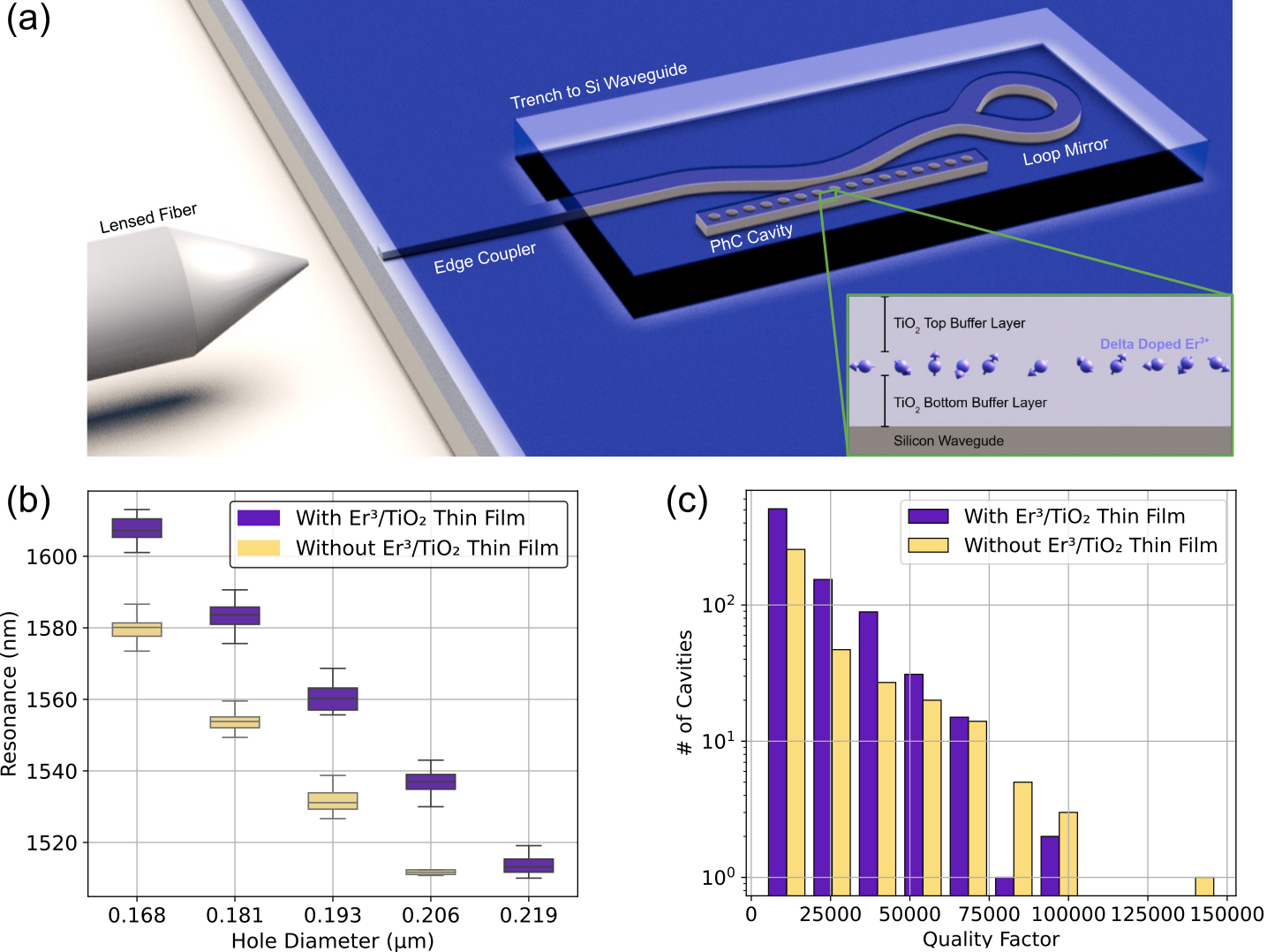} %Figure3
\caption{Portrayal of photonic crystal cavity with trench opening for deposition of quantum layer. The inset shows the cross section of the  Er$^{3+}$:TiO$_2$ layer (b) Box and whisker plot (interquartile range and +/- 1.5 quartiles) of the cavity resonance wavelength as a function of photonic crystal cavity hole diameter before and after deposition of the Er$^{3+}$:TiO$_2$ layer. These devices were from a different reticle than those showed in Figure 2. (c) Histogram of device's Q-factors before and after deposition of the Er$^{3+}$:TiO$_2$ layer. }
\end{figure*}
 
After room temperature characterization, the sample is placed in a closed–cycle cryostat at $T = 3.4$~K and a lensed optical fiber is aligned to the PIC edge couplers. We perform photoluminescence excitation (PLE) measurements and observe Er$^{3+}$ ensemble emission in both the rutile and anatase phases near 1520.5~nm and 1532.8~nm respectively, as shown in Fig. 4a. The linewidth of the rutile ensemble, which is the dominant phase present in the film and the focus of this study, is measured to be 43~GHz. The main contributions to the linewidth are expected to be charged defects such as oxygen vacancies \cite{singhOpticalMicrostructuralStudies2024}, though disorder from grain boundaries may also play a role. Ensemble emission from the ions is characterized in waveguides without resonant cavities. We observe optical lifetimes of 6.86(44)~ms for ions in rutile and 1.61(32)~ms for ions in anatase. Our results are consistent with previous investigations of Er$^{3+}$ ensembles in the anatase phase \cite{jiNanocavityMediatedPurcellEnhancement2024} and slightly longer than previous investigations of Er$^{3+}$ ensembles in the rutile phase \cite{phenicieNarrowOpticalLine2019,dibosPurcellEnhancementErbium2022,solomonAnomalousPurcellDecay2024}. In the remainder of this work we focus on the characterization of single ions in the rutile phase. Previous work has investigated single ions in anatase phase TiO$_2$, though not on a commercial scale foundry platform \cite{jiIsolationIndividualEr2024}. 
% We note that TiO$_2$ is birefringent and the longer lifetime observed here is consistent with the ordinary polarization axis of rutile TiO$_2$ \cite{phenicieNarrowOpticalLine2019}.

When solid-state quantum emitters are deposited atop the cavities, evanescent coupling of the optical dipoles to the cavity anti-node yields considerable reduction in the radiative lifetime through the Purcell effect \cite{purcellSpontaneousEmissionProbabilities1995}. Cavities with resonance wavelengths near 1520.5~nm are tuned through the rutile Er$^{3+}$:TiO$_2$ ensemble emission linewidth by controlled freezing of nitrogen gas onto the cavity \cite{mosorScanningPhotonicCrystal2005}, enabling the optical isolation of single ions. Gas condensation provides a convenient way to tune a single cavity's resonance; however, this process is incompatible in systems that would require multiple cavities operating simultaneously. Cavities made from electro-optic materials or post-fabrication index trimming are promising paths towards achieving multiple individually tuned cavities on a single chip \cite{jayatilleka_post-fabrication_2021, yang_controlling_2023}. 

All single ion measurements shown here are performed with an over coupled cavity that has a linewidth of $\kappa/2\pi = 5.28$~GHz and a quality factor of Q$\approx$37,400 shown in the Supporting Information. The efficiency of the edge coupler for the device was measured to be 50.3\%, in line with the design specification, and the overall photon detection efficiency of the setup is approximately 7\%, discussed in the Supporting Information. The Purcell enhanced optical lifetime, measured when the ion is resonant with the cavity, is shown in Fig. 4b. The enhanced optical lifetime is found to be $T_1$=13.80(56)~$\mu$s, giving a Purcell factor of $P=496(38)$ when compared with the emission lifetime of the rutile ensemble in the waveguide. We infer a single photon coupling rate of $g/2\pi=3.9$~MHz from the relationship $P=4g^{2}/(\kappa\Gamma_0)$, where $\Gamma_0=1/T_1$ of the ensemble outside the cavity. We observe single ion count rates for the ion presented here of $\approx$~170~Hz. We have observed single ion count rates of $\approx$~400~Hz on similar devices. We hypothesize that the observed count rates may be limited by spectral diffusion of the ion resonance with respect to the excitation laser \cite{gerardCrossoverInhomogeneousHomogeneous2025}.

The optical linewidth of a representative single ion, observed through resonant PLE, is shown in Figure 4c. The ion displays a full width at half maximum linewidth of 57.7~MHz in a single scan fit with a Voigt lineshape, larger than the radiative limit of 11.5~kHz ($\Delta\nu_{\mathrm{rad}}=1/(2\pi \: T_{1})$) for the $T_1$ observed here. The Voigt lineshape includes the parameters $\gamma$, $\sigma$ as the Lorentzian half width at half maximum and Gaussian standard deviation, respectively. The inhomogeneous contribution to the single scan linewidth is found to be $2\sqrt{2\ln{2}} \: \sigma=50.4$~MHz, while the homogeneous contribution is found to be $2\gamma=13.5$~MHz. Each single scan was recorded over approximately 5 minutes. We note here that the homogeneous contribution to the linewidth includes all diffusion and broadening mechanisms resulting in a Lorentzian spectrum, including for example the radiative linewidth, optical power broadening, instantaneous spectral diffusion, and pure dephasing. In this work we are not able to isolate individual contributions to the homogeneous linewidth. Over the course of approximately 2 hours, the ion displays additional slow spectral wandering as shown in the inset. The long--term linewidth over this time span is fully described by its inhomogeneous character and is well fit by a Gaussian lineshape, giving a full width at half maximum of $\Gamma_\mathrm{inhom}=78.8$~MHz.

The linewidths of single Er$^{3+}$ ions in the vicinity of nanophotonic devices are typically observed to be on the order of a few 10s of MHz as seen in YSO (5 MHz) \cite{dibosAtomicSourceSingle2018}, Si (70 MHz) \cite{gritschPurcellEnhancementSinglephoton2023b}, and lithium niobate (20-40 MHz) \cite{yang_controlling_2023}. Si nanophotonic cavities stamped on bulk CaWO$_4$ have recently demonstrated single ion linewidths of nearly 200~kHz \cite{ourariIndistinguishableTelecomBand2023b}, demonstrating that sub--MHz linewidths can be achieved in the vicinity of nanophotonic devices. We note that the films studied here have not yet been optimized in terms of growth and post-growth treatment to achieve narrower emitter linewidths, and we expect that fluctuations related to grain boundaries, surfaces, and point-like defects within the films all contribute to the observed broadening as observed in rare--earth doped nanocrystals \cite{bartholomewOpticalLineWidth2017,liuControlledSizeReduction2018}. A closely related avenue to explore is tailoring the thickness of the thin film buffering layers to reduce the proximity of the ions to interfaces as well as surface passivation techniques \cite{liuSingleSelfAssembled$mathrmInAs2018}. In addition to film growth and preparation, further improvements can be achieved to bring the observed optical linewidths closer to the radiative limit. Higher Purcell factor cavities can be used to shorten the optical lifetime directly, thereby increasing the radiative linewidth limit. For the results presented here we have not made use of the highest quality factor cavities available on the chip (see Fig. 3c), which were not within gas tuning range of the Er$^{3+}$ ensemble, and higher quality factor cavities should be possible with continued process improvements \cite{panuski_full_2022}.

\begin{figure*}
\centering\includegraphics[]{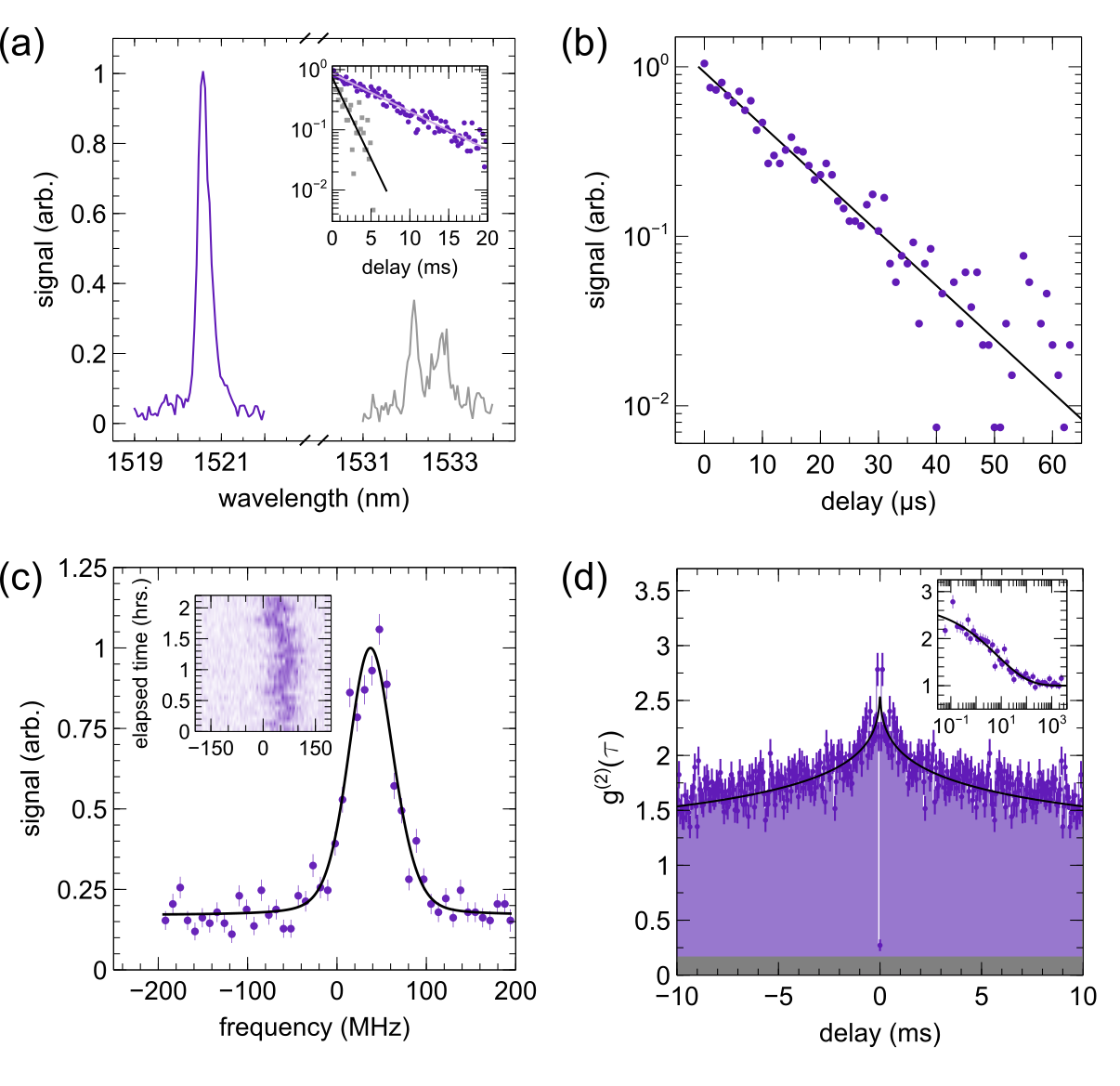} %width=12cm
\caption{(a) Photoluminescence excitation of the Er$^{3+}$:TiO$_2$ film showing the inhomogeneous emission from ions in the rutile (purple) and anatase (gray) phases of TiO$_2$. Inset: Optical lifetime of Er$^{3+}$ ions in each ensemble measured in an on-chip waveguide. (b) The Purcell enhanced optical lifetime of a single ion in the rutile phase in a resonant cavity. (c) Photoluminescence excitation scan across a single ion in rutile TiO$_2$ and fit with a Voigt lineshape with an inhomogeneous contribution of 50.4~MHz and a homogeneous contribution of 13.5~MHz to the total linewdith. Inset: Long-term observation of the single ion linewidth for a period $>2$~ hours, over which the inhomogeneous linewidth increases to 78.8~MHz. (d) Single photon autocorrelation indicating single photon emission from a single Er$^{3+}$ ion on a silicon nanobeam cavity. Inset: Bunching in the single photon autocorrelation viewed out to a timescale $>1$~s. }
\end{figure*}

Finally, we confirm single photon emission by measuring antibunching in the photon autocorrelation $g^{(2)}(\tau)$, shown in Fig. 4d. In the measurement we use a shot repetition period $\approx 4 T_1$ to reduce inter-shot correlations, with each data point recorded by integrating the total signal from each shot. The data are symmetric about zero delay because the measurement is performed on a single detector. At zero time delay we observe $g^{(2)}(0)=0.27(4)$, which indicates single photon emission for $g^{(2)}(0)<0.5$. If we take the independently measured background into account, shown in gray, we note that the zero delay autocorrelation improves to $g^{(2)}(0)=0.10$. In addition to the observed antibunching at zero delay, we also observe photon bunching for time delays $|\tau|>0$. In the inset we measure the autocorrelation out to time delays $\tau>1$~s, which is more than four orders of magnitude greater than the Purcell enhanced emitter lifetime. We see that the value of the autocorrelation returns to the Poissonian limit of $g^{(2)}(\tau)=1$ after about 100~ms. This bunching effect is characteristic of resonantly excited emitters undergoing spectral diffusion \cite{delteilPhotonStatisticsResonantly2024,sallenSubnanosecondSpectralDiffusion2010}. We find that the bunching correlation displays a non-exponential decay, which we characterize with a stretched exponential function to highlight the deviation from a purely exponential decay. We model the bunching for $|\tau|>0$ as $g^{(2)}(\tau)=B_{0}e^{-(\tau/ \tau_{B})^n}+1$, where $B_{0}$ is the bunching amplitude, $\tau_{B}$ is the characteristic bunching timescale, and $n$ is a stretching parameter that can take values $0<n<1$. Here we find a bunching amplitude of $B_{0}=1.66$, with a characteristic bunching time $\tau_{B}=7.3$~ms and a stretching parameter $n=0.38$. We discuss the form of this fit function in more detail in the Supporting Information. The degree of bunching can be used to estimate the homogeneous linewidth of an emitter relative to the inhomogeneous linewidth over the course of the measurement. In the case of a Gaussian broadening process, evidenced by the Gaussian inhomogeneous linewidth observed here ($\Gamma_\mathrm{inhom}=78.8$~MHz), the bunching amplitude when the emitter is excited on resonance can be inferred by \cite{delteilPhotonStatisticsResonantly2024}, 
$$B_{0}=\frac{1}{2\sqrt{\pi\ln{2}}}\frac{\Gamma_{\mathrm{inhom}}}{\Gamma_{\mathrm{hom}}}-1,$$

\noindent
where it has been assumed that the inhomogeneous linewidth is much greater than the homogeneous linewidth for simplicity. From the observed bunching amplitude, we can therefore estimate the homogeneous linewidth to be $\Gamma_\mathrm{hom}=10.0$~MHz, which is in reasonable agreement with the estimate determined from PLE scans shown in Fig 4c. 

%\section{Conclusion}
In conclusion, we have demonstrated silicon photonic crystal cavities on AIM Photonics' Quantum Flex 300~mm platform. We further demonstrated backend deposition of Er$^{3+}$:TiO$_2$ thin films, which offers a scalable approach to developing single-photon emitters with spin-photon interfaces in the telecom C--band. Furthermore, Er$^{3+}$ has been previously demonstrated as an exceptional quantum memory, making this platform ideal for future development of quantum networking systems such as interconnects and quantum repeaters. The results presented here are from post-processing after the wafer left the foundry, although we expect the quality of the roughness and crystallinity to be improved if the material deposition is part of the foundry’s process flow, enabling a fully integrated quantum platform. Beyond quantum networking, this work demonstrates that subwavelength photonic devices can be readily fabricated at the foundry level, paving the way for a range of new technologies. \\

\begin{acknowledgement}
The authors thank Alan Dibos and F.J. Heremans for helpful discussions. Authors R.M.P., A.S., S.G., M.K.S., and S.E.S. acknowledge support from the U.S. Department of Energy, Office of Science, Advanced Scientific Computing Research (ASCR) program under Grant No. CRADA A22112 through the Chain Reaction Innovations program at Argonne National Laboratory. Work performed at the Center for Nanoscale Materials, a U.S. Department of Energy Office of Science User Facility, was supported by the U.S. DOE, Office of Basic Energy Sciences, under Contract No. DE-AC02-06CH11357. This material is based on research sponsored by the Air Force Research Laboratory under AIM Photonics (agreement number FA8650-21-2-1000).  The U.S. Government is authorized to reproduce and distribute reprints for Governmental purposes notwithstanding any copyright notation thereon. The views and conclusions contained herein are those of the authors and should not be interpreted as necessarily representing the official policies or endorsements, either expressed or implied, of the United States Air Force, the Air Force Research Laboratory or the U.S. Government.
\end{acknowledgement}

\begin{suppinfo}
 Supporting Information: Waferscale variation of photonic crystal cavities, Bragg filters, silicon nitride photonic crystal nanobeam cavities, cryogenic single ion characterization. This material is available free of charge via the internet at http://pubs.acs.org.
\end{suppinfo}

%%%%%%%%%%%%%%%%%%%%%%% References %%%%%%%%%%%%%%%%%%%%%%%%%
\bibliography{Foundry_Bib}

\begin{figure}[h]
\centering
\includegraphics[]{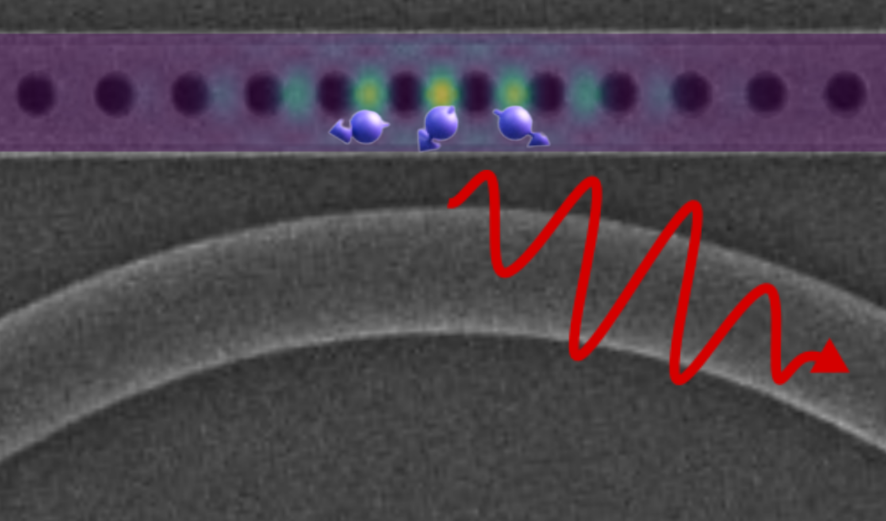}
\caption*{For Table of Contents Only}
\end{figure}

\end{document}